\documentclass[journal]{IEEEtran}
\ifCLASSINFOpdf
\else
\fi

\usepackage[colorlinks, citecolor=blue,bookmarks=false]{hyperref}
\usepackage{booktabs}
\usepackage{setspace, amsmath, amssymb, url, lscape, subfig, algorithmic, multirow, pslatex, listings, verbatim, alltt, amsfonts, wrapfig, boxedminipage, color, cite, pifont} 
\usepackage{caption} 
\usepackage[vlined,linesnumbered,ruled,boxed]{algorithm2e}
\usepackage[dvips]{graphicx}
\usepackage{epsfig}
\usepackage{dblfloatfix}
\usepackage{xcolor}
\usepackage{balance}

\newcommand{\qed}{\nobreak \ifvmode \relax \else
      \ifdim\lastskip<1.5em \hskip-\lastskip
     \hskip1.5em plus0em minus0.5em \fi \nobreak
      \vrule height0.75em width0.5em depth0.25em\fi}

\newcommand{\paperfolder}{.}

\newcommand{\eg}{{\it e.g., }}
\newcommand{\etal}{{\it et~al., }}
\newcommand{\ie}{{\it i.e., }}

\newcommand{\comments}[1]{}
\newcommand\hl{\bgroup\markoverwith
  {\textcolor{yellow}{\rule[-.5ex]{2pt}{2.5ex}}}\ULon}

\newlength{\boxfigwidth}
\newcommand{\boxfig}[1]{
\begin{figure}[h]
\begin{center}
\begin{small}
\setlength{\boxfigwidth}{3.15in}
\addtolength{\boxfigwidth}{0in}
\noindent\framebox{\quad\begin{minipage}{\boxfigwidth}
#1
\vspace{-15pt}
\end{minipage}\quad}
\end{small}
\end{center}
\end{figure}
}

\begin{document}
\title{Improving Robustness of Heterogeneous Serverless Computing Systems Via Probabilistic Task Pruning}

\author{%
  Chavit Denninnart, *James Gentry, Mohsen Amini Salehi
\\ High Performance Cloud Computing (HPCC) Laboratory,\\ School of Computing and Informatics, University of Louisiana at Lafayette, USA 
\\  \{cxd9974,  amini\}@louisiana.edu, *gentry@hpcclab.org
}
\maketitle              
\begin{abstract}
Cloud-based serverless computing is an increasingly popular computing paradigm. In this paradigm, different services have diverse computing requirements that justify deploying an inconsistently Heterogeneous Computing (HC) system to efficiently process them.
In an inconsistently HC system, each task needed for a given service, potentially exhibits different execution times on each type of machine. An ideal resource allocation system must be aware of such uncertainties in execution times and be robust against them, so that Quality of Service (QoS) requirements of users are met. 
This research aims to maximize the robustness of an HC system utilized to offer a serverless computing system, particularly when the system is oversubscribed. 
Our strategy to maximize robustness is to develop a task pruning mechanism that can be added to existing task-mapping heuristics without altering them. Pruning tasks with a low probability of meeting their deadlines improves the likelihood of other tasks meeting their deadlines, thereby increasing system robustness and overall QoS.
To evaluate the impact of the pruning mechanism, we examine it on various configurations of heterogeneous and homogeneous computing systems. Evaluation results indicate a considerable improvement (up to 35\%) in the system robustness. 

\end{abstract}


\section{Introduction}
\label{sec:intro}

The cloud-based serverless computing paradigm abstracts away the details of resource provisioning and deployment~\cite{baldini2017serverless}. A user only needs to define the required services (aka functions or tasks) and their QoS expectations within the serverless platform. 
The user is only charged for the actual resource usage and not for the idle times or under-utilized provisioned resources~\cite{eivy2017wary}. Because of its ease-of-use and cost-efficiency, serverless platforms have gained popularity over the past few years, specifically for micro-service deployments~\cite{baldini2017serverless, sill2016design}.

As a general computing system, on the back-end, the serverless platform receives service requests from multiple users. Each service request implies executing one or more tasks that potentially have diverse computing and memory demands. This diversity justifies the use of Heterogeneous Computing (HC) systems to improve both QoS satisfaction and incurred cost~\cite{baldini2017serverless, yan2016building}. Heterogeneity of an HC systems can be divided into qualitative and quantitative differences. Differences in architecture can be described as qualitative (\eg GPU-based versus CPU-based architecture) whereas differences in performance within a given architecture can be described as quantitative (\eg one computer has a faster CPU speed than the other). A system with both qualitative and quantitative differences between machines is described as \emph{inconsistently heterogeneous} while a system with only quantitative difference is described as consistently heterogeneous~\cite{ali2000,Maheswaran97,Braun01,li2018cost}. 

Arriving service requests (also, termed \emph{tasks}) can also be qualitatively and quantitatively heterogeneous. A qualitative heterogeneity refers to different types of tasks in the workload (\eg compressing a video segment versus changing its spatial resolution~\cite{li2018cost}) whereas quantitative heterogeneity can refer to characteristics such as variety in data size within a certain task type (\eg compressing video segments with different lengths). Qualitative heterogeneity among tasks leads to differences in execution times on different machine types in an HC system~\cite{Shestak08}, known as \emph{task-machine affinity}~\cite{li2018cost}. For instance, embarrassingly parallel tasks (\eg image filtering) perform faster on (\ie have higher affinity with) GPU-based machines whereas data-intensive tasks with many branching paths have higher affinity on CPU-based machines with large memory~\cite{Shestak08}. Quantitative heterogeneity within a given task type causes variation (uncertainty) in the execution time of tasks of that type for each machine type within the HC system. In an inconsistently heterogeneous system, making optimal mapping of arriving tasks on HC systems is desirable, but not practically feasible~\cite{coffman76,Ibarra77} owing to the high level of uncertainty and a large decision space.


\begin{figure*}%
    \centering 
    \subfloat[\small{Immediate-mode resource allocation.}]{{\includegraphics[width=0.24\textwidth]{\paperfolder/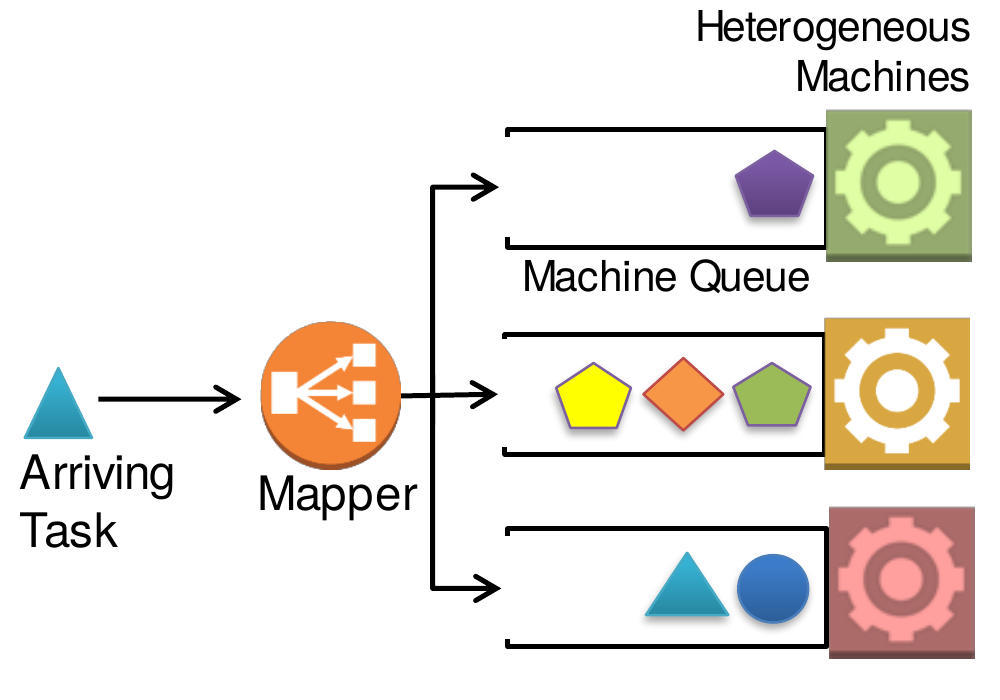} }  
    \label{fig:ImmSchedule}
    }
    \hspace{0.5cm}%
    \subfloat[\small{Batch-mode resource allocation.}]{{\includegraphics[width=0.312\textwidth]{\paperfolder/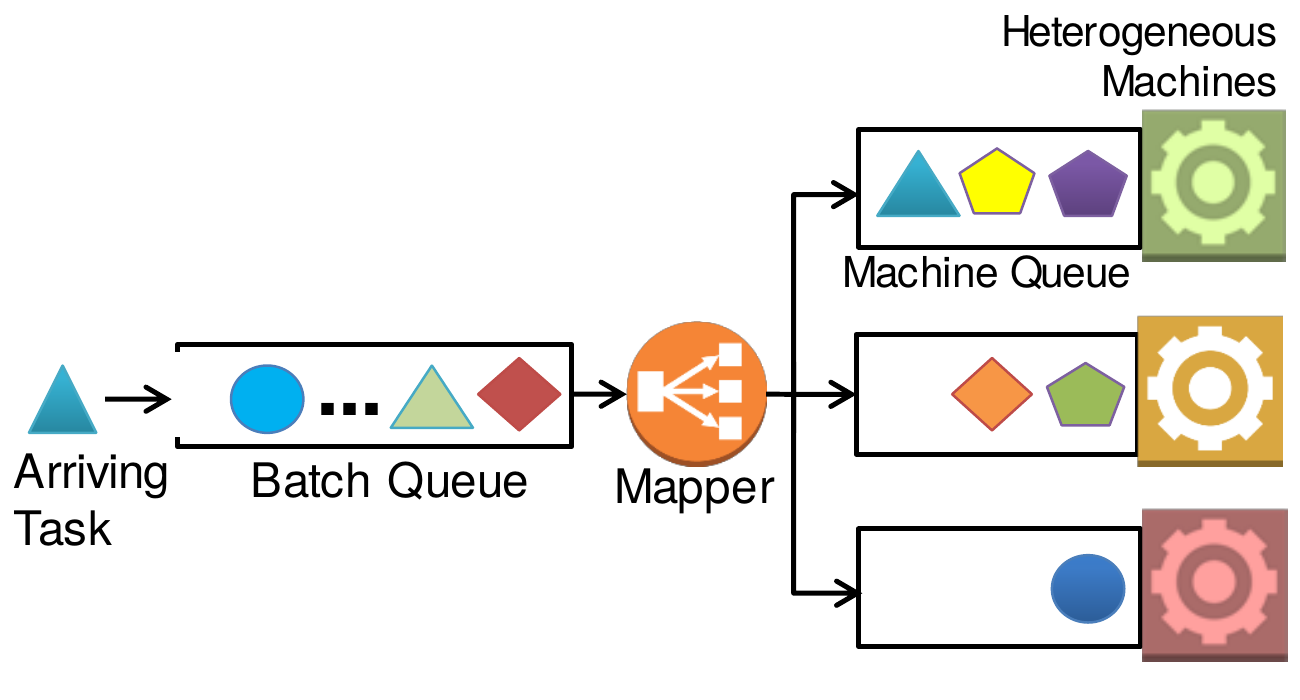} }
    \label{fig:BatchSchedule}
    }
    \hspace{0.5cm}%
    \subfloat[\small{Batch-mode resource allocation, with task pruning mechanism attached.}]
    {{\includegraphics[width=0.312\textwidth]{\paperfolder/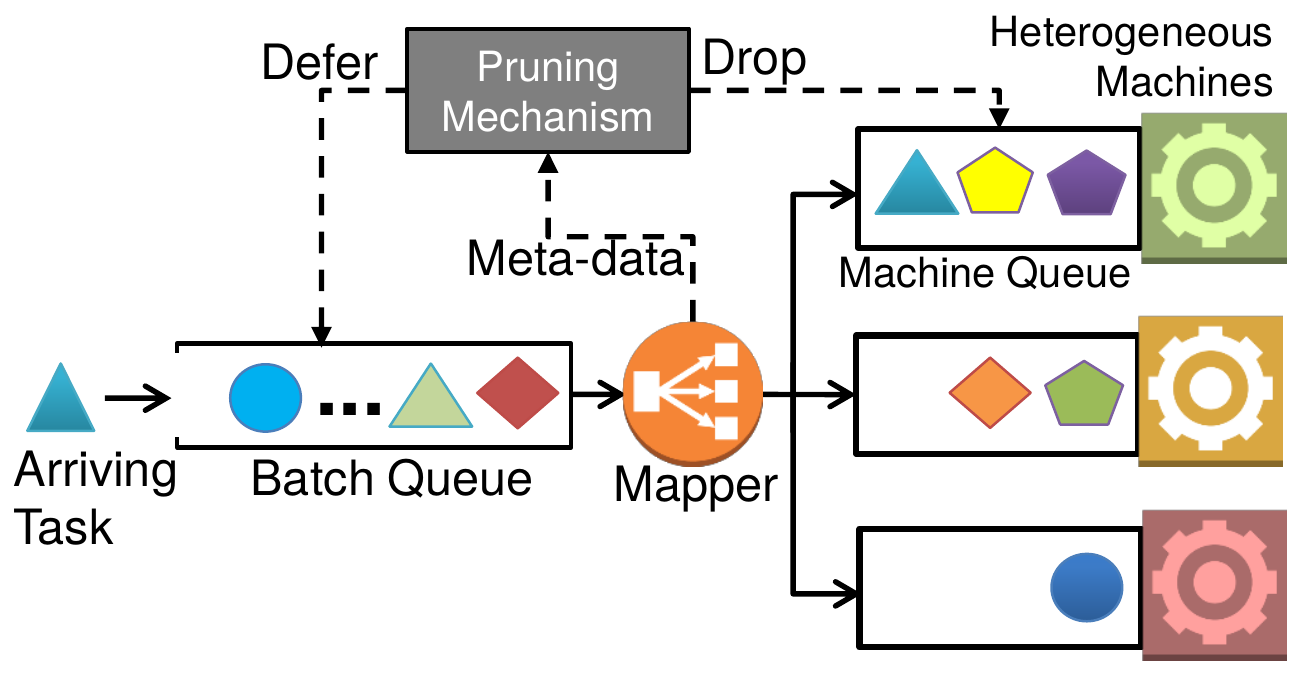} }
    \label{fig:BatchSchedulePrune}
    }
    \caption{\small{Batch-mode resource allocation operates on batches of tasks upon task completion (and task arrival when machines queues are not full). Immediate-mode resource allocation operates immediately, upon arrival of a task to the system. Pruning mechanism can be plugged into existing resource allocation system. }}    

    \label{fig:Scheduler}%
\end{figure*}

In this study, we consider the case where each arriving task has an individual hard deadline that has to be met to fulfill the task's QoS requirement. That is, there is no value in executing a task after its deadline, therefore, a task that is past its deadline must be dropped from the system~\cite{khemka2015utility,Khemka2014UDD, khemka14utility}. As an example, live video streaming tasks that miss their presentation times (\ie deadlines) must be dropped~\cite{li2016vlsc} to catch up with the current live video contents. As tasks' deadline violations often occur when the system is heavily loaded, we particularly study circumstances in which the HC system is \emph{oversubscribed}. That is, the intensity of arriving workload is such that it is impossible to finish all the tasks by their deadlines~\cite{salehi2016stochastic}.

\emph{Robustness} of a system is defined as its ability to maintain a given level of \underline{performance} in the face of \underline{uncertainty}~\cite{salehi2016stochastic}. The main \emph{goal} in this study is to maximize the robustness of an inconsistently HC system used for processing tasks of a serverless computing platform. The performance metric to measure robustness of the HC system is based on the number of tasks completing before their deadlines. To achieve the robustness, we harness two sources of uncertainty that exist in HC systems, namely uncertainty in tasks' execution times and uncertainty in tasks' arrival rate~\cite{TPDS17,chen2015towards}. 

To attain the robustness goal, our hypothesis is that the resource allocation of an HC system should not map tasks that are unlikely to  complete on time (\ie before its individual deadline). In fact, such mappings just increase the incurred cost~\cite{salehi10} without improving the robustness of HC system. More importantly, executing unlikely-to-succeed tasks postpones the execution of other pending tasks, that subsequently miss their deadlines. Therefore, our strategy is to avoid executing unlikely-to-succeed tasks. The existing resource allocation systems of an HC system in a serverless computing platform operate either in the \emph{immediate}- or \emph{batch}- modes. In the former, as shown in Figure~\ref{fig:Scheduler}\subref{fig:ImmSchedule}, the mapper makes mapping decision immediately upon task arrival. In the latter, as shown in  Figure~\ref{fig:Scheduler}\subref{fig:BatchSchedule}, arriving tasks are batched in a queue awaiting mapping to a machine at a later mapping event. Although both of these approaches are being utilized, prior studies suggest that batch-mode mapping yields to a higher overall robustness~\cite{salehi2016stochastic,li2018cost, malone1983enterprise}.

We design a \emph{pruning mechanism} that is plugged into the resource allocation system (as shown in Figure~\ref{fig:Scheduler}\subref{fig:BatchSchedulePrune}) and improves the robustness of the HC system. The mechanism receives tasks' meta-data 
 from the mapper and prune unlikely-to-succeed tasks. The pruned tasks are either deferred to a later mapping event or dropped to alleviate the oversubscription of the HC system.

Assume that there are a few pending tasks in a machine queue awaiting execution ahead of an arriving task. Each one of the pending tasks has uncertainty in its execution time; collectively, they introduce compound uncertainty for the arriving task's execution time. We need to calculate the impact of this compound uncertainty on the likelihood of success for an arriving task. Then, the pruning mechanism should determine if the arriving task is worth assigning to the machine. In an HC system, for an arriving task that has a low chance of success (\ie chance of meeting its deadline), deferring its assignment to the next mapping event can potentially increase its chance of success. This is because such deferment provides the opportunity for the arriving task to be assigned to a machine with a higher affinity that may become available in a later mapping event. Furthermore, when the system is heavily loaded (\ie oversubscribed), it is possible to take an even more aggressive approach and drop (\ie discard) pending tasks with low chance of success to increase the likelihood that other pending tasks succeed. 

We note that, making task mapping decisions only based on chance of success can potentially impact fairness of the system. In this case, the pruner consistently favors tasks with short execution time, while pruning other longer tasks. To avoid this problem, we equip the pruning mechanism with a method to achieve fairness across various types of tasks exist in an HC system. The advantage of our proposed pruning mechanism is that it does not require making any change in the existing resource allocation and mapping heuristic of the systems. Instead, the pruning mechanism is plugged into the existing mapping heuristic that best suites to the system and improves its robustness. 

To evaluate the proposed pruning mechanism, we plug it into various widely-used mapping heuristics \cite{madni2017performance, ezzatti2013efficient} that are used in batch- and immediate-mode resource allocation systems. In addition to HC systems, we study the impact of applying pruning mechanism on mapping heuristics of resource allocations in homogeneous computing systems. 

In summary, the contributions of this paper are as follows:
\begin{itemize}
 \item Proposing a probabilistic task pruning mechanism within resource allocation system to enable robust heterogeneous serverless computing.
 \item Extending pruning mechanism to enable fairness across various task types in the system. 
 \item Plugging the pruning mechanism into popular mapping heuristics used in resource allocation systems.
  \item Analyzing the impact of task pruning mechanisms on the robustness of various types of computing systems and under varying workload characteristics.
\end{itemize}

Extensive simulation results on various workloads demonstrate that our proposed mechanism increases robustness remarkably (by up to 35 percentage point). More importantly, the impact of this pruning mechanism is more substantial in HC systems that are highly overloaded.

\begin{figure*}[hb!]
  \centering
  \includegraphics[width=0.6\textwidth]{\paperfolder/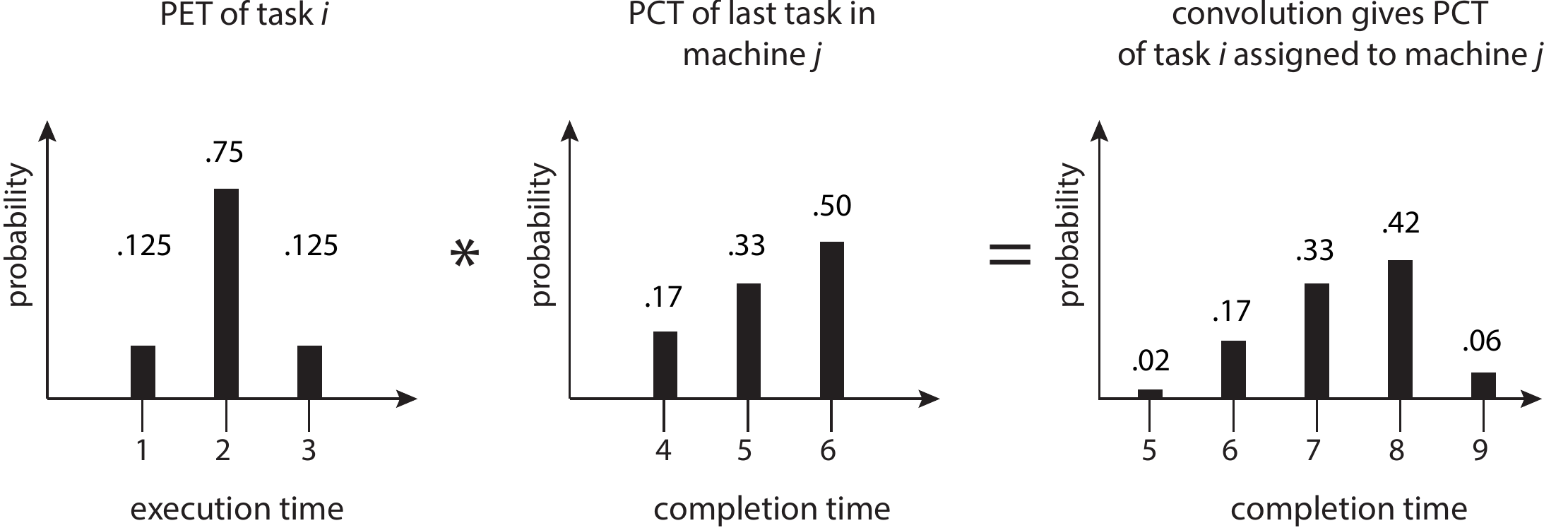}
  \caption{Probabilistic Execution Time (PET) of an arriving task is convolved with the Probabilistic Completion Time (PCT) of the last task on machine $j$ to form the PCT for the arriving task $i$ on the assigned machine. \label{fig:conv} }
\end{figure*}

The rest of this paper is organized as follows. 
Section~\ref{sec:probstate} establishes the problem statement and provides mathematical background required in this study.
Then, Section \ref{sec:background} introduces commonly used mapping heuristics. 
Next, Section \ref{sec:pruning} details task pruning mechanism and 
Section \ref{sec:evltn} describes and analyzes evaluation results. 
In Section \ref{sec:rw}, we highlight some prior works related to this study.
Finally, Section \ref{sec:conclsn} concludes the paper and offers direction for potential future works.

\section{System Model}
\label{sec:probstate}

This research is motivated by a serverless computing platform with heterogeneous machines, such as those used in edge computing~\cite{nastic2017serverless}. Another motivation is a cloud-based serverless platform where the service provider has budget constraints~\cite{salehi10,wu2012sla}. In these scenarios, the service provider aims to maximize the number of service requests meeting their deadline constraint within their limited resources.
Users issue independent service requests (termed \emph{tasks}) from a set of offered service types (termed \emph{task-types}). 
A task in our study is modeled as an independent video segment in the form of Group Of Pictures (GOPs) that is sequentially processed (\eg transcoded~\cite{TPDS17}) within a deadline constraint. 
Each task has an individual hard deadline, which is the presentation time of that video segment~\cite{icsoc18,li2016vlsc}. As there is no value in executing a task that has missed its deadline, the task must be dropped from the system.

In the described scenario, both inconsistent task heterogeneity, in form of different task-types (\eg transcoding types), and inconsistent machine heterogeneity, in form of different machine types, can be present. In serverless computing, mapping tasks to machines is hidden from the user and is carried out using mapping methods of the resource allocation system. 

In our system model, tasks dynamically arrive into the resource allocation system. Their arrival rate and pattern is unknown in advance. In particular, we are interested to study robustness of the system when there is a surge in demand. We define \emph{oversubscription} as a situation in which task arrival is so intense that it is impossible to complete all tasks within their deadlines. 

Due to variations in tasks' data sizes, the execution time of each task type on each machine type is stochastic~\cite{Shestak08}. For instance, the execution time to change the resolution depends on the size of the GOP to be processed~\cite{TPDS17}. The stochastic execution time of each task type on each machine type is modeled as a Probability Mass Function (PMF)~\cite{salehi2016stochastic}. In an inconsistently HC system, a Probabilistic Execution Time (PET) matrix is used to represent execution time distribution of each task type on each machine type~\cite{Shestak08}. For an arriving task, based on the PET matrix, we can calculate its Probabilistic Completion Time (PCT) distribution on a given machine. As shown in Equation~\ref{eq:pct}, the PCT of arriving task $i$ on machine $j$, denoted $PCT(i,j)$, is calculated by convolving PET of task $i$ on machine $j$, denoted $PMF(i,j)$, with PCT of the last task already mapped to machine $j$, denoted $PCT(i-1,j)$.
 \begin{equation}\label{eq:pct}
  PCT(i,j) = PMF(i,j) * PCT(i-1,j)
 \end{equation} 
 
Figure~\ref{fig:conv} shows an example for calculating PCT of arriving task $i$ on a given machine $j$ based on its PET on this machine convolved with PCT of the last task already assigned to machine queue $j$. Once we calculate PCT for arriving task $i$ on machine $j$, we can obtain its \emph{chance of success} (denoted as $S(i,j)$), which is defined as the probability that task $i$ completes before its deadline. Equation~\ref{eq:0} formally represents the chance of success for task $i$ with deadline $\delta_i$ on a given machine $j$.

\begin{equation} \label{eq:0}
	S(i,j) = \mathbb{P}(PCT(i,j) \leq \delta_i)
\end{equation}
 
Since PCT of task $i$ relies on the PCT of tasks ahead of it in the machine queue, as the queue length grows, the compound uncertainty in task's completion time grows too. Conversely, when the queue length is shortened because of task dropping, the PET of the dropped task is no longer used in the convolution process to calculate PCT of tasks behind the dropped one. Hence, their PCT is changed (as explained in~\cite{salehi2016stochastic,ipdps19}) in a way that their compound uncertainty is reduced and the chance of success for the affected tasks is increased.

Our proposed pruning mechanism operates in coordination with a mapping method at each mapping event and decides to either drop or defer tasks with a low chance of success. A mapping event occurs when a task completes its execution or when a new task arrives into the system. Before any mapping decision is confirmed, the system drops any task that has missed its deadline. Due to the overhead of data transfer, we assume that a task cannot be remapped once it is mapped to a machine. After mapping, tasks assigned to a machine queue are processed in a First Come First Serve manner. Each task is executed in isolation on the machine without preemption or multitasking~\cite{dogan04,cao13}.

\section{Mapping Heuristics}
\label{sec:background}

\subsection{Overview}
\begin{figure*} 
  \centering
  \includegraphics[width=0.6\textwidth]{\paperfolder/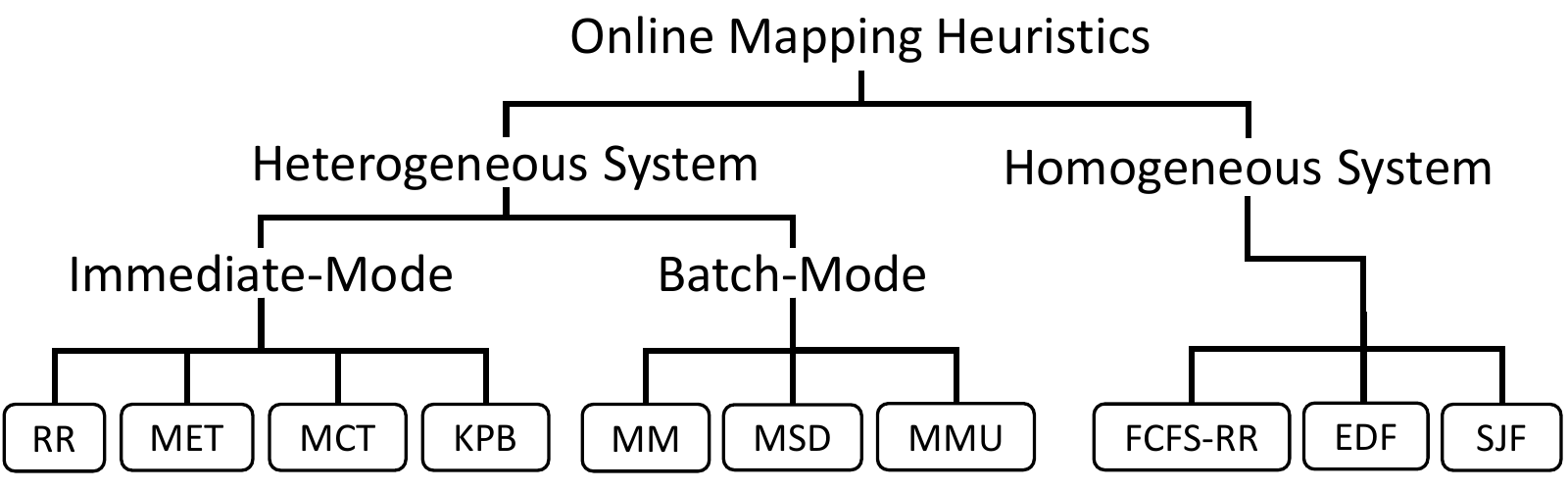}
  \caption{Overview of mapping heuristics widely-used in various types of distributed systems. \label{fig:convtree} }
  \vspace{-5px}
\end{figure*}

Although this study concentrates on the performance of pruning in HC systems, to conduct a comprehensive analysis, in addition to HC systems, we study the performance of pruning both on homogeneous systems as well.
In both types of systems, optimal task-machine mapping is proven to be an NP-complete problem \cite{coffman76,Ibarra77}. Therefore, a large body of mapping heuristics (\eg~\cite{Braun01,Ibarra77,madni2017performance}) have been developed for these systems. Figure~\ref{fig:convtree} provides an overview of mapping heuristics commonly used in heterogeneous and homogeneous systems. In particular, mapping heuristics of HC systems can be further categorized based on those operate in immediate-mode and in batch-mode resource allocation systems. 

Immediate-mode mapping heuristics do not hold tasks in an arrival queue and they are simpler to develop. In batch-mode heuristics, however, mapping occurs both upon task arrival (when machine queues are not full) and upon task completion. 
Batch-mode heuristics generally use an auxiliary structure, known as \emph{virtual queue}, where arriving tasks are examined on different machine queues. These heuristics commonly use a two-phase process for decision making. In the first phase, the heuristic finds the best machine for each task, by virtue of a per-heuristic objective. In the second phase, from task-machine pairs obtained in the first phase, the heuristic attempts to choose the best machine-task pairs for each available machine queue slot. After all slots are filled, or when the unmapped queue is emptied, the virtual mappings are pushed (assigned) to the machine queues, and the mapping method is complete.

Although mapping heuristics used in homogeneous computing systems are of batch nature, their logic is simpler than those used in batch-mode of HC systems. In the rest of this section, we review widely-used heuristics in both heterogeneous and homogeneous computing systems.

\subsection{Immediate-Mode Mapping Heuristics for Heterogeneous Systems}

\paragraph{Round Robin (RR)}  
In RR~\cite{li2016high}, incoming tasks are assigned in a round robin manner to an available machine, from Machine 0 to Machine $n$.

\paragraph{Minimum Expected Execution Time (MET)} 
In MET, the incoming task $i$ is assigned to the machine that offers the minimum expected execution time (\ie the average of the $PET(i,j)$ for task $i$ on machine $j$). 

\paragraph{Minimum Expected Completion Time (MCT)} In MCT, the incoming task is assigned to the machine that offers the minimum expected completion time. The completion time is obtained based on the accumulated expected execution time of tasks queued in a given machine.

\paragraph{K-Percent Best (KPB)} 
KPB is a combination of MCT and MET. It only considers MCT amongst the $K$ percent of machines with the lowest expected execution times for an incoming task.


\subsection{Batch-Mode Mapping Heuristics for Heterogeneous Systems}

\paragraph{MinCompletion-MinCompletion (MM)} 
MM is a popular mapping heuristic widely-used in the literature~\cite{he2003qos,pedemonte2016accelerating,ezzatti2013efficient, salehi2016stochastic}. The PET matrix is used to calculate expected completion times. In the first phase of this two-phase heuristic, the virtual queue is traversed, and for each task in that queue, the machine with the minimum expected completion time is found, and a pair is made. In the second phase, for each machine with a slot in its machine queue, the provisional mapping pairs are examined to find the machine-task pair with the minimum completion time, and the assignment is made to the virtual machine queues. The process repeats itself until all virtual machine queues are full, or until the unmapped queue is exhausted.

\paragraph{MinCompletion-Soonest Deadline (MSD)}
MSD is a two-phase process, first selecting the machine which provides the minimum expected completion time (using the PET matrix) for each task under consideration. In second phase, from this list of possible machine-task pairs, the tasks for each machine with the soonest deadline are chosen, and in the event of a tie, the task with the minimum expected completion time breaks the tie. 
As with MM, the process is repeated until either the virtual machine queues are full, or the unmapped task queue is empty.

\paragraph{MinCompletion-MaxUrgency (MMU)}
MMU is a two-phase process. The first phase is identical to MM and MSD. The second phase select task based on their urgency. Urgency for task $i$ on machine $j$ is defined as the inverse of the difference between the task deadline ($\delta_i$) and the expected completion time of the task on machine $j$ ($\mathbf{E}[C(t_{ij})]$). Equation~\ref{eq:urg} formally shows the urgency definition.

\begin{equation}\label{eq:urg}
    U_{ij} = \frac{1}{\delta_i-\mathbf{E}[C(t_{ij})]}
\end{equation}

As with MM and MSD, this process is repeated until either the temporary batch queue is empty, or until the virtual machine queues are full.
%


\subsection{Mapping Heuristics for Homogeneous Systems}

\paragraph{First Come First Served - Round Robin (FCFS-RR)}~
In FCFS-RR, a task is selected in first come first serve manner and is assigned to the first available machine in a round robin manner, from Machine 0 to Machine $n$. 

\paragraph{Earliest Deadline First (EDF)}~
EDF is functionally similar to MSD heuristic for HC systems. The first phase finds the machine with the least expected completion time. Then, the second phase sorts the arrival queue in an ascending order based on tasks' deadlines. Next, the task in the head of the arrival queue is assigned to the machine with minimum expected completion time. This process is repeated until all task are mapped or the machine queues are full.

\paragraph{Shortest Job First (SJF)}~
SJF is functionally similar to MM heuristic for HC systems. The first phase finds the machine with the least expected completion time. Then, the second phase sorts the arrival queue in an ascending order based on tasks' expected execution time. Next, the task in the head of the arrival queue is assigned to the machine with minimum expected completion time. This process is repeated until all task are mapped or the machine queues are full.

\section{Pruning Mechanism}
\label{sec:pruning}

\subsection{Overview}
\begin{figure} 
  \centering
  \includegraphics[width=0.50\textwidth]{\paperfolder/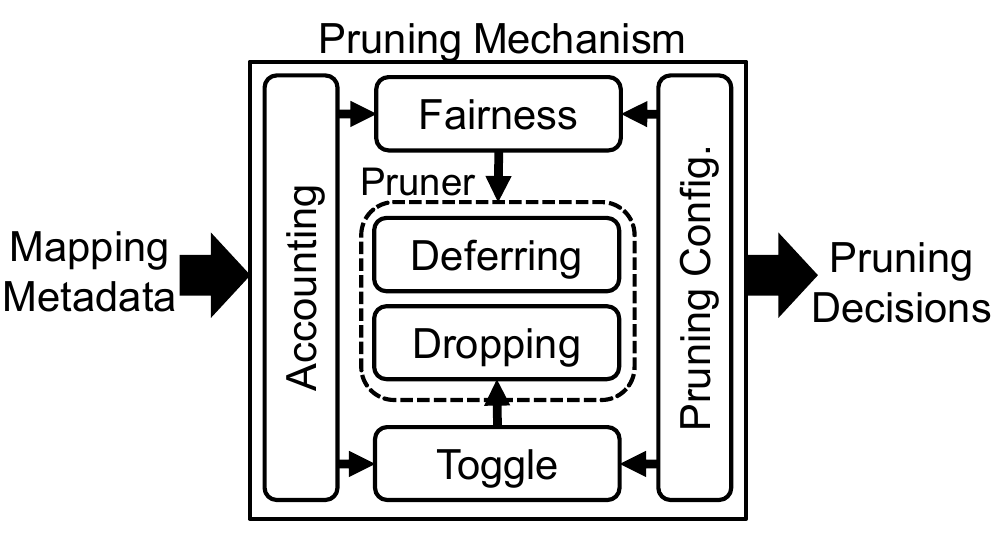}
  \caption{\small{Architectural overview of the task pruning mechanism. Pruning Configuration and Accounting provide inputs to Toggle and Fairness components. These inputs determine the Pruner decisions.} \label{fig:PruneArch} }
\end{figure}

In this section, the probabilistic task pruning mechanism (also termed Pruner) is described. Figure~\ref{fig:PruneArch} shows a pruning mechanism that can be plugged into any resource allocation system and work in conjunction with any mapping heuristic to increase the robustness of the HC system. The \emph{Accounting} module is in charge of gathering tasks' information (\ie meta-data) from the resource allocation system.
The \emph{Toggle} module measures the oversubscription level of the HC system based on the collected information and decides whether the task dropping has to be engaged. Aiming at maximizing robustness, the \emph{Pruner} module, first, calculates the chance of success for all tasks, then, enacts the pruning decision on tasks whose chance is lower than a user-defined threshold, specified in \emph{Pruning Configuration}. 
The \emph{Fairness} module keeps track of the suffered task types (\ie those that are consistently dropped) and adjust the Pruner to avoid biasness against them. 

In the rest of this section, we first elaborate on each of these modules and their interactions. Then, we explain the overall pruning procedure.

\subsection{Task Deferring}
Within a mapping event, task deferring is defined as the act of postponing assigning a task, whose chance of success is low, to a machine. In an HC system, such deferment can be beneficial both to the deferred task and to the overall system robustness. For the deferred task, it is possible that its likelihood of success increases over the next mapping event(s), as a machine that offers a lower completion time (\ie a higher chance of success) can potentially become available. In addition, deferring unlikely-to-succeed tasks can possibly improve the chance of success for other waiting tasks which ultimately enhances the overall system robustness. We note that, task deferring is applied in every mapping event and only on tasks waiting in the arrival queue. 

Effectively, task deferring limits mapping heuristics to only map tasks whose chance of success is higher than a user-defined (\ie service provider) threshold, called \emph{Pruning Threshold}. By setting the threshold, a service provider can manage his/her incurred cost of using cloud-based serverless platform wasted to process tasks that are unlikely to succeed. 


\subsection{Task Dropping}
Task dropping is defined as the act of evicting a task from the system. Task dropping can occur in two manners, namely reactive and proactive dropping. Reactive task dropping occurs when a task has already missed its deadline. 
Proactive task dropping, however, predictively drops tasks whose chance of success is low, before their deadline is reached. Proactive dropping is considered a more aggressive pruning decision and should be enacted only under high levels of oversubscription. Such task dropping does not only increase the chance of success for the tasks behind the dropped one, but also reduces the compound uncertainty in their completion time, which yields more informed/accurate mapping decision. Hence, task dropping in a sufficiently oversubscribed system potentially improves the overall system robustness.

Toggle module is in charge of determining when the system is sufficiently oversubscribed to shift to a more aggressive pruning via triggering task dropping. The current implementation of Toggle checks the number of tasks missing their deadlines since the previous mapping event and identifies the system as oversubscribed if the number is beyond a configurable \emph{Dropping Toggle}.
 
\subsection{Fairness Preservation}
Aiming at maximizing robustness, pruning can get biased to task types with shorter expected execution time, because such tasks generally provide a higher chance of success. The pruning unfairness can be harmful, as it consistently violates deadlines of certain task types.
To avoid such a bias, the Accounting module keeps track of tasks that are dropped or completed. Then, the Fairness module utilizes the tasks' information to calculate the \emph{sufferage score} of each task type.

Each on-time completion of a task of type $k$ reduces the sufferage score of task type $k$ by a configurable constant value ($c$), whereas dropping a task of type $k$ increases the sufferage score of its type by $c$ value. The constant value $c$, termed \emph{fairness factor}, can be adjusted to determine how aggressively the sufferage score changes. The sufferage score of task type $k$ is then used as a probability offset to the Pruning Threshold of such task type.




\subsection{Pruning Procedure}
\label{subsec:prunproc}
Detailed procedure of the pruning mechanism is explained in form of a pseudo-code in Figure~\ref{fig:algdrop}.
At the beginning of each mapping event, tasks that have already missed their deadlines are dropped (Step 1).
The Fairness module uses the information collected by the Accounting module about tasks completed since the previous mapping event to update the sufferage score (Step 2).      
If the the system is identified as oversubscribed by the Toggle module (Step 3), the Pruner is engaged to drop tasks that are unlikely to succeed from machine queues. For that purpose, the chance of success for each task is calculated (Step 5) and those whose chance is less than the Pruning Threshold are dropped (Step 6). After mapping decisions are made (and before dispatching tasks to machines), the pruner defers assigning tasks whose chance of success is less than the Pruning Threshold ($\beta$) to the next mapping event (Step 10). The remaining tasks are dispatched to their assigned machines (Step 11). Steps 7---11 are repeated until there is no remaining task in the batch queue or machine queues are full.

\boxfig{
$\alpha \leftarrow$ Dropping Toggle

$\beta \leftarrow$ Pruning Threshold

$c \leftarrow$  Fairness Factor

$\gamma_{1..n} \leftarrow$ Fairness Score of all task types


\bigbreak

Upon triggering of a mapping event: 
\begin{itemize}
\item[(1)] Drop all pending tasks that missed their deadlines
\item[(2)] Collect and process data of all task $t_f$ completed on-time since previous mapping event
\begin{itemize}
	\item[--] $k \leftarrow $ task type of $t_f$ 	
	\item[--] $\gamma_k \leftarrow \gamma_k - c$
\end{itemize}  
  
\item[(3)] If oversubscription level is greater than $\alpha$
\begin{itemize}
	\item[(4)] For each task $t_i$ of type $k$ on machine queue $m_j$:
		\begin{itemize}
			\item[(5)] Calculate $chance(i,j)$
			\item[(6)] If $chance(i,j) \leq \beta - \gamma_k$
			\begin{itemize}
			\item[--] Drop $t_i$
			\item[--] $\gamma_k \leftarrow \gamma_k + c$
			\end{itemize}
		\end{itemize}
\end{itemize}

\item[(7)] While there is unmapped task and machine queues are not full:
\begin{itemize}
	\item[(8)] Call mapping heuristic
	\item[(9)] For each task $t_i$  of type $k$ mapped to machine $j$:
	\begin{itemize}

		\item[(10)] If $chance(i,j) \leq \beta - \gamma_k$
		\begin{itemize}
			\item[--] Defer $t_i$ to the next mapping event
		\end{itemize}
	\end{itemize}
\item[(11)] Dispatch remaining assigned tasks to machines
\end{itemize}
\end{itemize}

\caption{\small{Pruning mechanism procedure at each mapping event.}}

 \label{fig:algdrop}
}

\section{Performance Evaluation}
\label{sec:evltn}

\subsection{Experimental Setup}
To evaluate the impact of pruning mechanism on a variety of widely-used mapping heuristics, we conducted a simulation study under various configurations of heterogeneous (in both immediate- and batch-modes) and homogeneous computing systems. For the experiments, Pruning Configurations are set to use Pruning Threshold of 50\% and Fairness factor of 0.05, unless otherwise stated. To accurately analyze the impact of dropping and deferring, we evaluate them both individually and together. 

For each set of experiments, 30 workload trials were performed using different task arrival times built from the same arrival rate and pattern. In each case, the mean and 95\% confidence interval of the results are reported. The experiments were performed using the Louisiana Optical Network Infrastructure (LONI) Queen Bee 2 HPC system~\cite{LONI}.  

%
While the task completion time estimation involves multiple convolutions which impose calculation overhead, there are multiple implementation techniques that can minimize the overhead of repeated calculation, such as task grouping and memorization of partial results.
Moreover, all the task pruning decisions are made by a dedicated machine which reserved for resource allocation. Therefore, pruning mechanism does not add extra overhead to each HC resources in our experiments.


\subsection{Workload Generation}

Twelve SPECint benchmarks were run numerous times on a set of eight machines which were used to generate probabilistic execution time (PET) PMFs~\cite{salehi2016stochastic}. The PMFs were generated by creating a histogram on a sampling of 500 points from a Gamma distribution formed using one of the means, and a shape randomly chosen from the range [1:20]. This was done for each of the twelve benchmarks, on each of the eight machines\footnote{The 8 machines are: Dell Precision 380 3 GHz Pentium Extreme, Apple iMac 2 GHz Intel Core Duo, Apple XServe 2 GHz Intel Core Duo, IBM System X 3455 AMD Opteron 2347, Shuttle SN25P AMD Athlon 64 FX-60, IBM System P 570 4.7 GHz, SunFire 3800, and IBM BladeCenter HS21XM.}, resulting in the eight by twelve machine type to task type PET matrix. The PET matrix remains constant across all of our experiments. 

In each experiment, a determined number of tasks per time unit is fed to the system within a finite time span. For each experiment, the system starts and ends in an idle state. As such, The first and last 100 tasks in each workload trial are removed from the data to focus the results on the portion of the time span where the system is oversubscribed. 

 \begin{figure} [h]
  \centering
  \includegraphics[width=0.31\textwidth]{\paperfolder/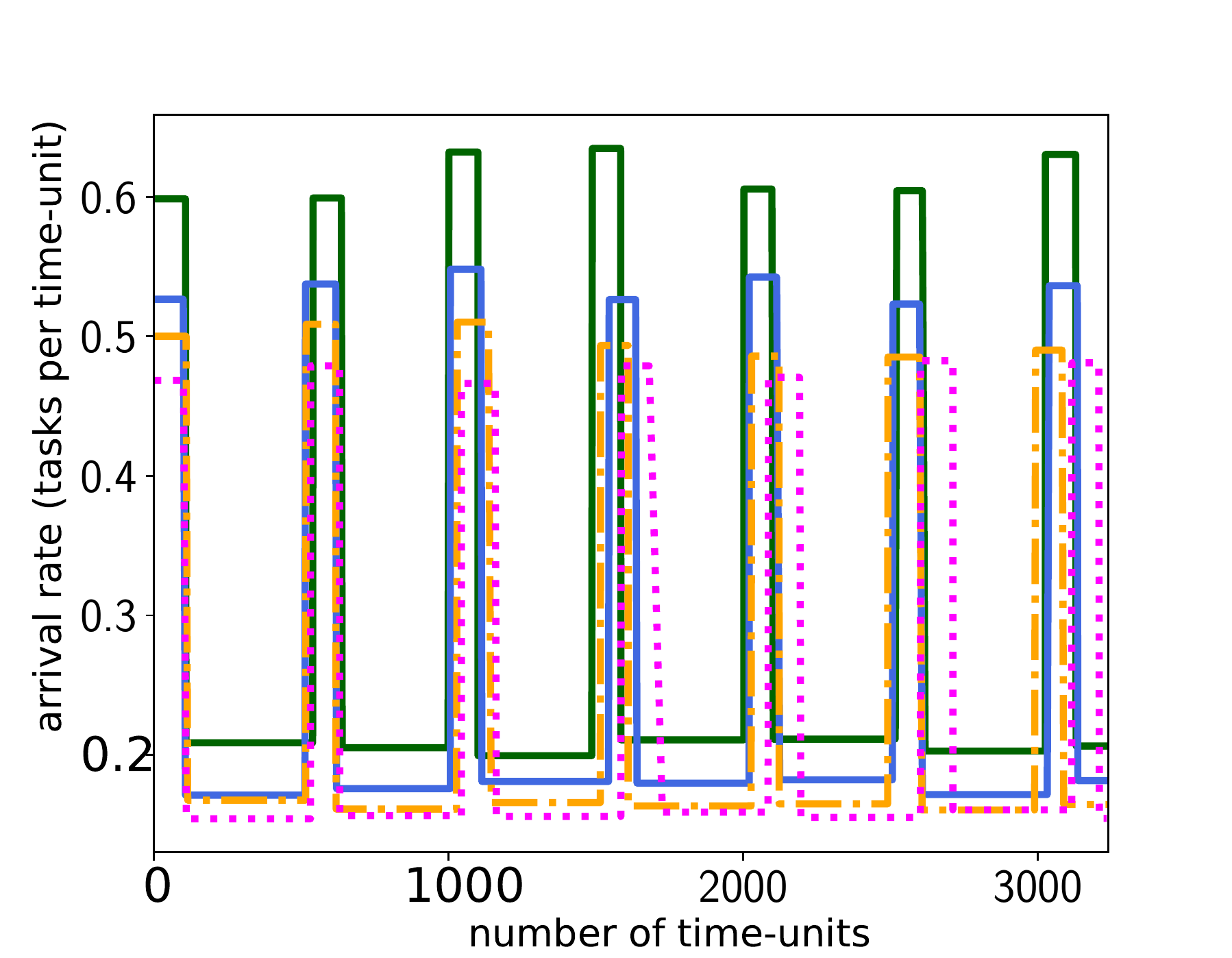}
  \caption{\small{Spiky task arrival pattern used in the experiments. Each color represents one task type. For better presentation, only four task types are shown. The Vertical axis shows the task arrival rate and horizontal axis shows the time span.} \label{fig:arrivalPattern} }
  \vspace{-7px}
\end{figure}

   \begin{figure*}[ht!]
    \centering 
    \subfloat[\small{Immediate-mode mapping heuristics}]{{\includegraphics[width=0.3\textwidth]{\paperfolder/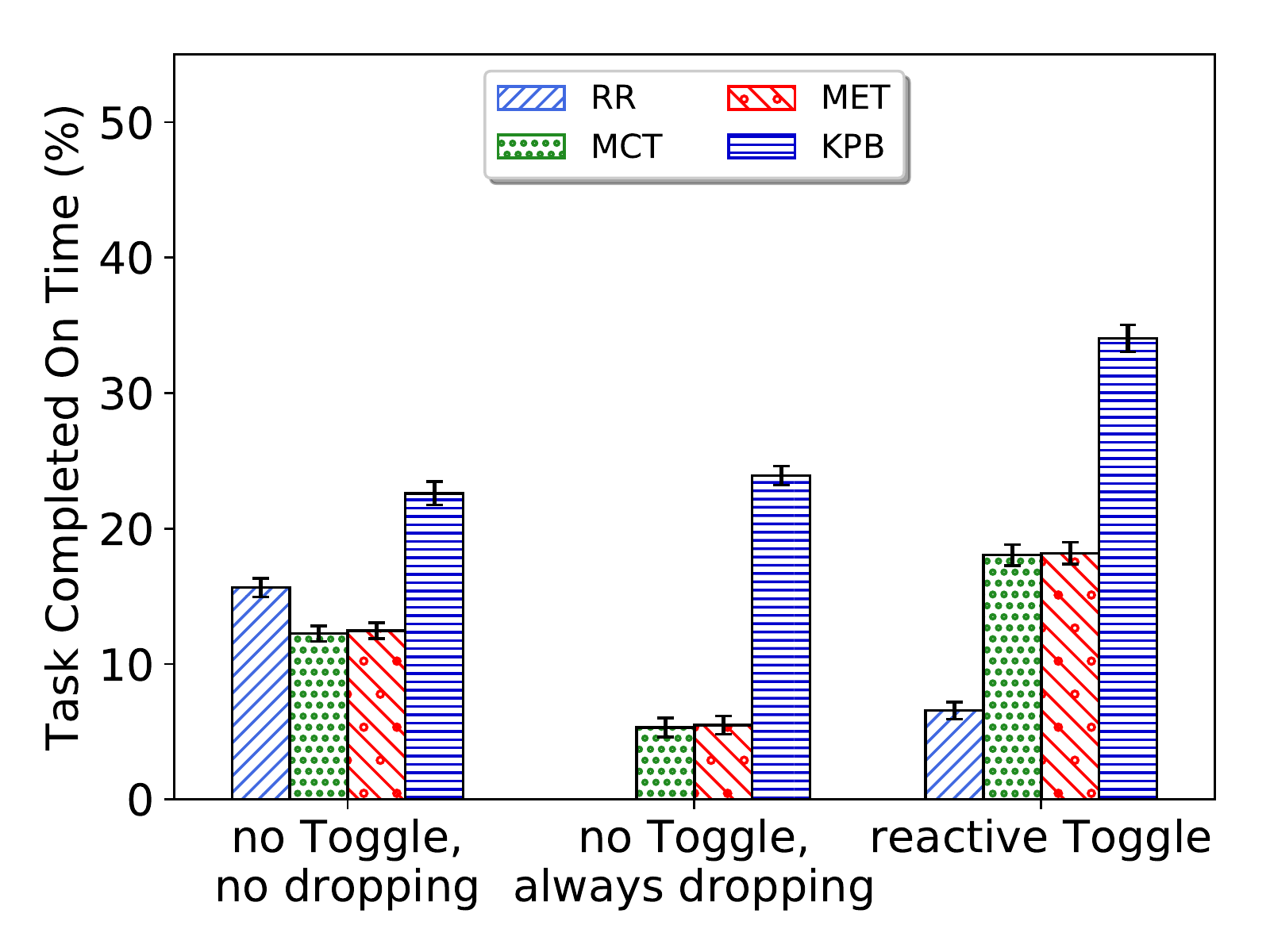} }
    \label{fig:arrivalImmediate}
    }
    \hspace{3.4cm}%
    \subfloat[\small{Batch-mode mapping heuristics}]{{\includegraphics[width=0.3\textwidth]{\paperfolder/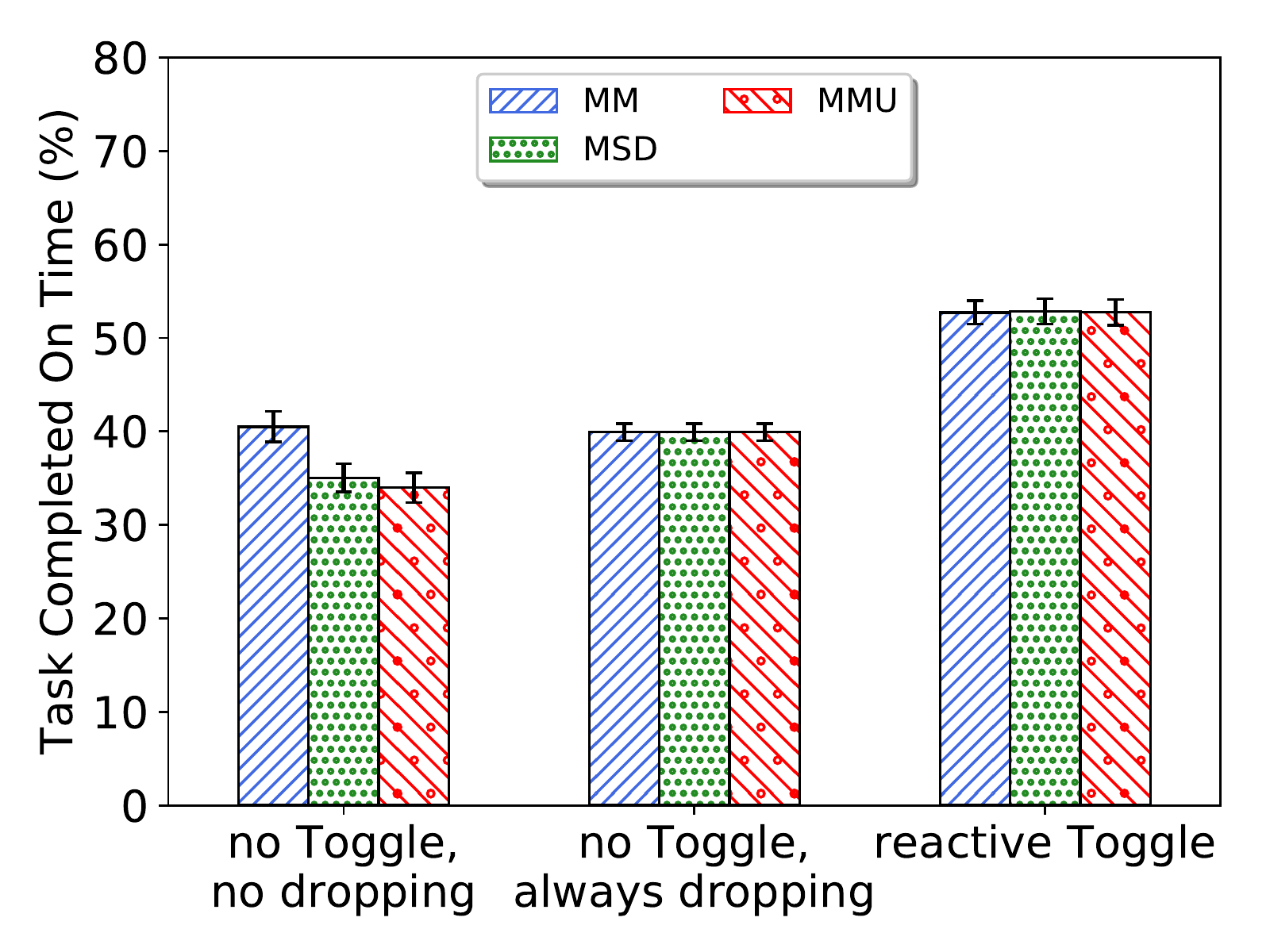} }  
    \label{fig:dropBatch}
	}
    \caption{\small{Impact of employing the Toggle module in a pruning mechanism works with immediate- and batch-mode heuristics. Horizontal axis shows how task dropping is engaged in reaction to oversubscription and vertical axis shows percentage of tasks completed on time. }}  
    \label{fig:toggle}%
    \vspace{-9px}
\end{figure*}

To conduct a comprehensive evaluation, two sets of workload were examined: \textbf{(A)} \emph{Constant rate arrival pattern}: a Gamma distribution is created with a mean arrival rate for all task types. The variance of this distribution is 10\% of the mean. Each task type's mean arrival rate is generated by dividing the number of time units by the estimated number of tasks of that type. A list of tasks with attendant types, arrivals times, and deadlines is generated by sampling from each task type's distribution. \textbf{(B)} \emph{Variable rate (spiky) arrival pattern}: In this case, tasks arrive with variable rates, as shown in Figure~\ref{fig:arrivalPattern}, to mimic arrival patterns observed in HC systems (\eg~\cite{miranda}). The spike times were determined uniformly, from the constant arrival workload, by dividing the workload time span to the number of spikes we want to create. During each spike, task arrival rate rises up to three times more than the base (lull) period. Each spike lasts for one third of the lull period. Since the spiky arrival pattern is frequently observed in real systems, it is our default workload arrival pattern in the experiments. 
For each task, as noted in Equation~\ref{eq:dl}, the deadline is calculated by adding the mean duration for that task type (\(avg_i\)) to the arrival time (\(arr_i\)), and then adding in a slack period based on the mean of all task type's duration multiplied by a tuning parameter (\(\beta \cdotp avg_{all}\)). This slack allows for the tasks to have a chance of completion in an oversubscribed system. In the workload trials, the value of $\beta$ of each task is randomly chosen from the range of $[0.8 , 2.5]$. 

\begin{equation}\label{eq:dl}
    \vspace{-3px}
\delta_i = arr_i + avg_i + (\beta \cdotp avg_{all})
\end{equation}

We carried out experiments under a variety of task arrival rates (oversubscription levels), however, the default rate used for plotting graphs includes 15K tasks that represents a moderately oversubscribed system. All the workload trials are publicly available from \url{git.io/fhSZW} for reproducing purposes.


\subsection{Impact of Toggle Reacting to Oversubscription in HC Systems}
In this experiment, our goal is to evaluate the impact of Toggle module within the pruning mechanism. Recall that the Toggle module is in charge of triggering task dropping operation. As such, we evaluate three scenarios: First, when there is no Toggle module in place and dropping operation is never engaged (referred to as ``no Toggle, no dropping"); Second, when Toggle module is not in place and task dropping is always engaged (referred to as ``no Toggle, always dropping"); Third, when the Toggle module is in place and is aware of (\ie reactive to) oversubscription (referred to as ``reactive Toggle"). In this case, the Toggle module engages task dropping only in observation of at least one task missing its deadline, since the previous mapping event. 

Figure~\ref{fig:arrivalImmediate} shows the results for the immediate-mode mapping heuristics and Figure~\ref{fig:dropBatch} shows them for the batch-mode. In both cases, we can observe that when Toggle functions in reaction to oversubscription, the overall system robustness is improved, regardless of the mapping heuristic deployed. The only exception is RR immediate-mode heuristic. The reason is that RR does not take execution time or completion time into account and it continuously maps tasks with a relatively low chance of success. These mapped tasks are subjected to be removed by task dropping. Without probabilistic task dropping, some of those low-chance tasks can complete on time.
 We can also observe that in immediate-mode, KPB provides the highest robustness (percentage of tasks completing on time) and also benefits the most from task dropping. This is because it makes more informed mapping decisions after dropping underperforming tasks.

The experiment testifies that our hypothesis in removing tasks with low chance of success in favor of  other tasks is true and can significantly improves robustness---by up to 12\% in immediate-mode and 19\% in batch-mode. 

\begin{figure}[ht]
  \centering
  \includegraphics[width=0.32\textwidth]{\paperfolder/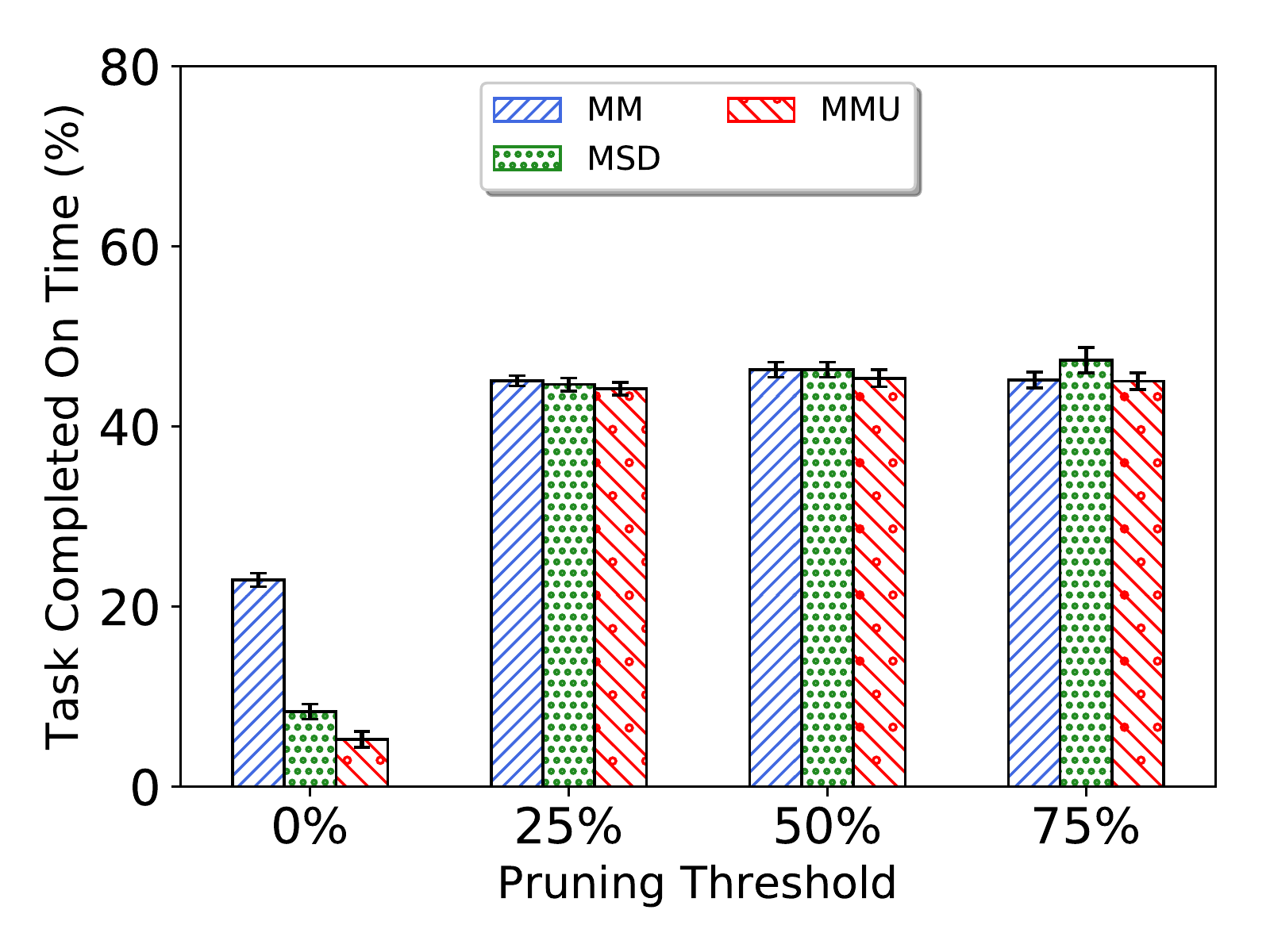}
  \caption{\small{Impact of tasks deferring on batch-mode mapping heuristics in an HC System with workload intensity of 25K. The Vertical axis shows percentage of tasks completed on time. The horizontal axis is the minimum success probability needed for each task to be mapped.   } \label{fig:deferBatch} }
  \vspace{-15px}
\end{figure}
   \begin{figure*}[hb!] %
    \subfloat[\small{Constant Arrival Pattern}]{{\includegraphics[width=0.29\textwidth]{\paperfolder/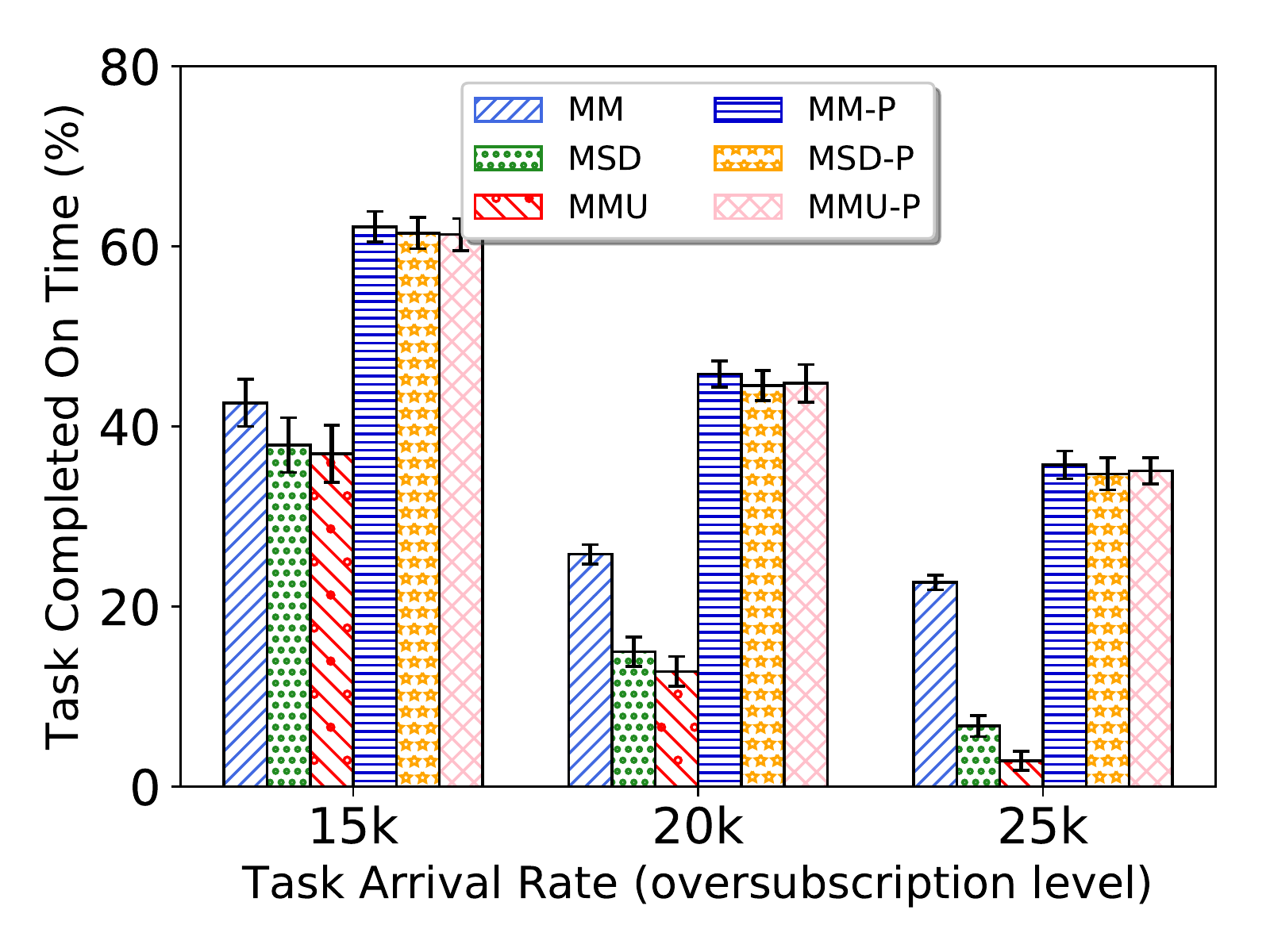} }  
    \label{fig:pruningBatch_c}
	}
    \hspace{3.4cm}%
	\centering 
    \subfloat[\small{Spiky Arrival Pattern}]{{\includegraphics[width=0.29\textwidth]{\paperfolder/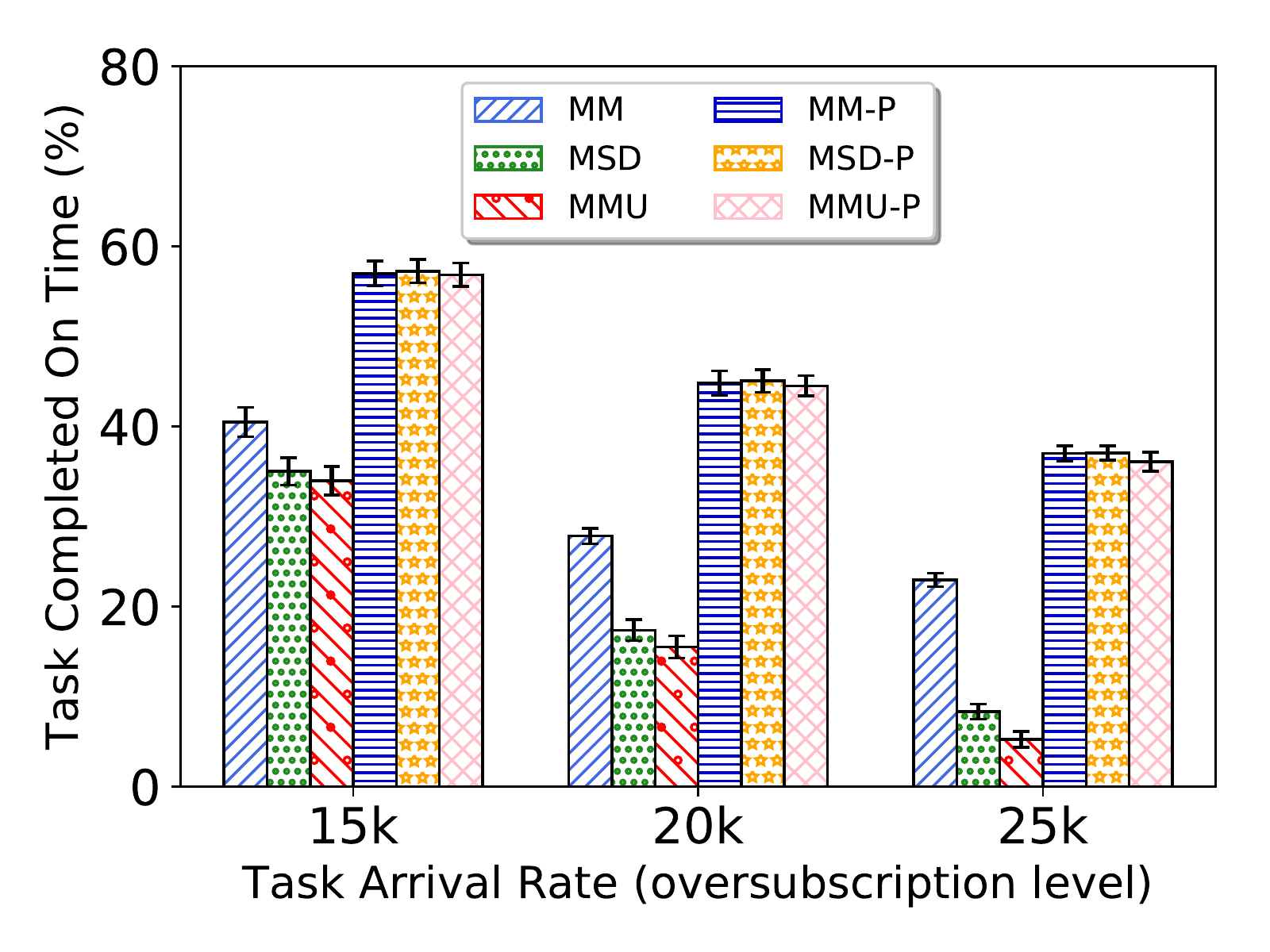} }
    \label{fig:pruningBatch_s}
    }
    \caption{\small{Impact of pruning mechanism on batch-mode heuristics in HC systems. Horizontal axes show the number of tasks arriving within a time unit (\ie oversubscription level). In the legend, ``-P" denotes heuristics use pruning mechanism.
    }}   
    \label{fig:pruningBatch}%
\end{figure*}

   \begin{figure*}[ht!]%
    \centering 
    \subfloat[\small{Constant Arrival Pattern}]{{\includegraphics[width=0.3\textwidth]{\paperfolder/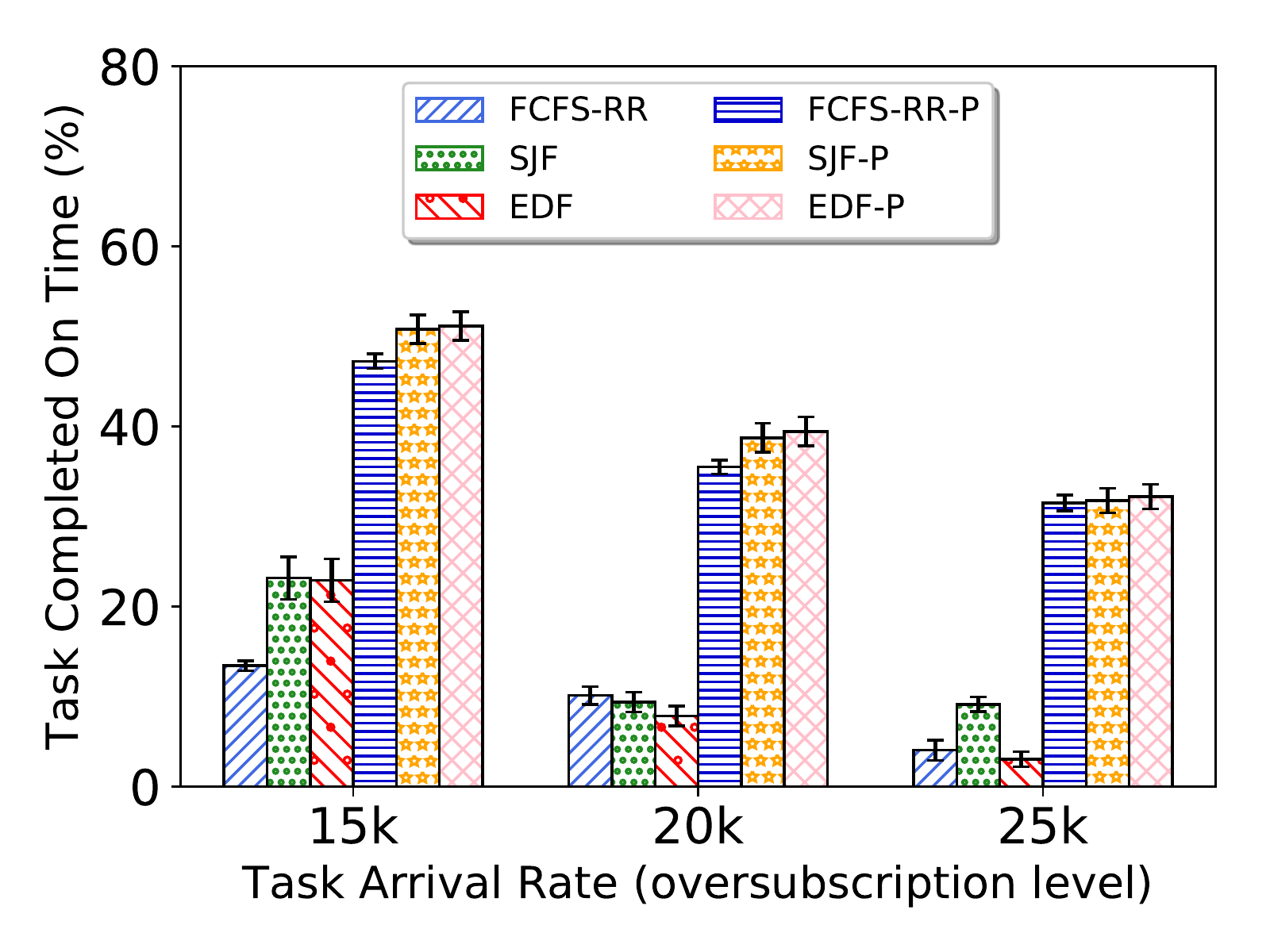} }  
    \label{fig:arrivalHomo_c}
	}
    \hspace{3.4cm}%
    \subfloat[\small{Spiky Arrival Pattern}]{{\includegraphics[width=0.3\textwidth]{\paperfolder/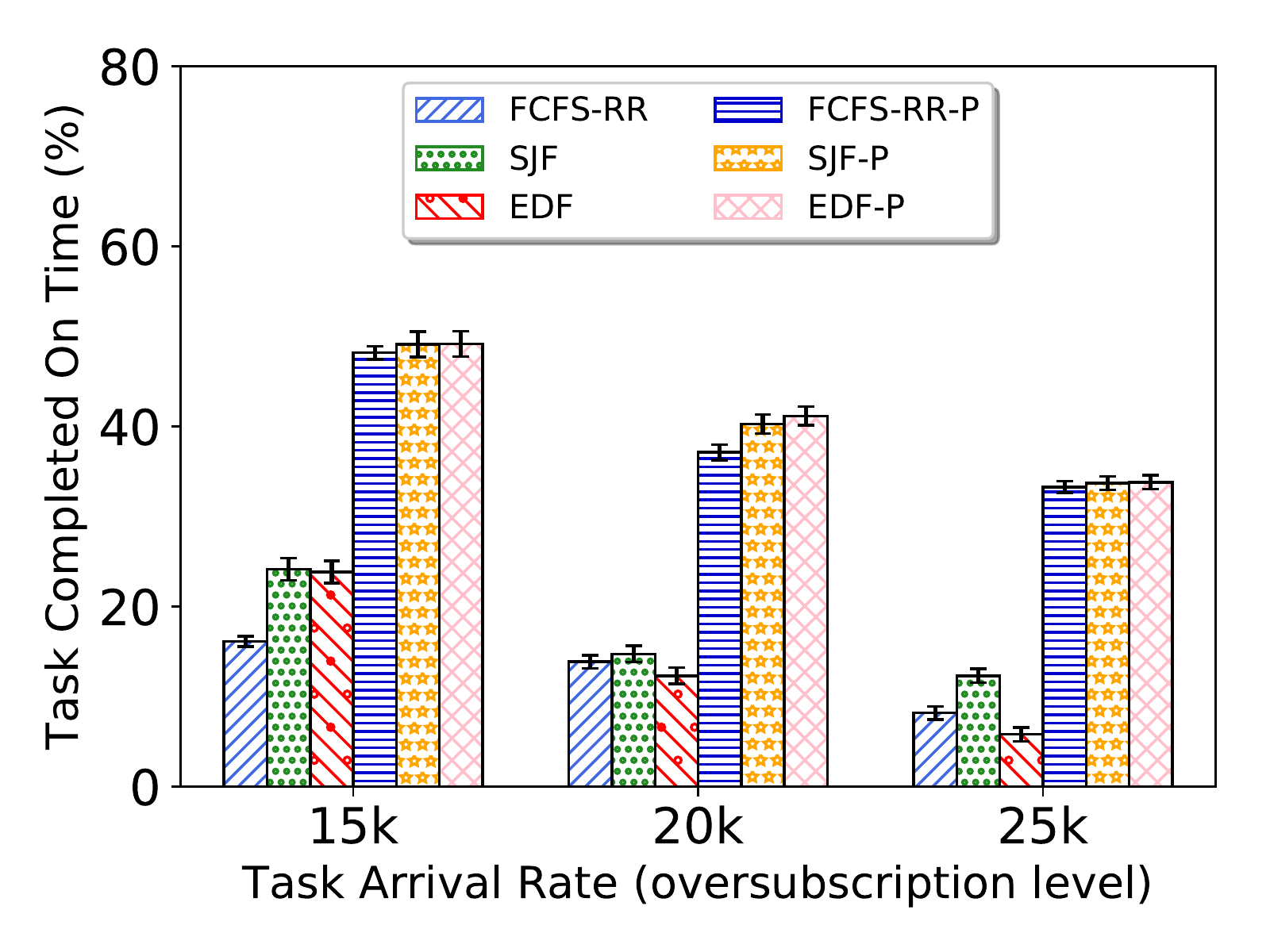} }
    \label{fig:arrivalHomo_s}
    }
    \caption{\small{Impact of pruning mechanism on mapping heuristics of homogeneous systems. Horizontal axes show the number of tasks arriving within a time unit (\ie oversubscription level). In the legend, ``-P" denotes heuristics use pruning mechanism.
    }}    
    \label{fig:arrivalHomo}%
    \vspace{-7px}
\end{figure*}

\subsection{Impact of Task Deferring on Batch-Mode Heuristics}
In this experiment, we evaluate the impact of task deferment within the pruning mechanism. As deferring operation works on the arrival (batch) queue, it can only be enabled for batch-mode heuristics. We conducted the experiment for task pruning threshold set to 0\% (no task pruning), 25\% , 50\%, and 75\%. As the results of this experiment is more prominent under high level of oversubscription, we set the task arrival to 25K tasks in the workload trials.

Figure \ref{fig:deferBatch} shows that, without task deferring (\ie when Pruning Threshold is zero), MM, MSD, and MMU's robustness are the lowest (between 5\% to 23\%). However, as task deferring is employed, all mapping heuristics can attain more than 44\% robustness. This is because pruning mechanism delays mapping of tasks with low chance of success until a more suitable machine becomes available. Hence, machines are utilized only to execute promising tasks, thereby increasing the robustness. 
 Our observation also implies that, for these widely used batch-mode mapping heuristics, by limiting the selection-pool of a mapping-heuristic to likely-to-succeed tasks, task deferring can reduce the performance differences of the heuristics to offer similar robustness, regardless of their algorithmic logic.


In Figure \ref{fig:deferBatch}, we can see that, in all heuristics, the robustness does not improve for Pruning Thresholds higher than 50\%. In fact, a high Pruning Threshold makes the system conservative in allocating tasks and defers tasks whose completion can improve overall robustness. Therefore, setting Pruning Threshold to 50\% is a proper configuration for the pruning mechanism.

\subsection{Impact of Pruning Mechanism on Batch-Mode Heuristics}
In this experiment, our goal is to evaluate the impact of the pruning mechanism holistically under various oversubscription levels. We evaluated the system robustness when mapping heuristics are coupled with and without the pruning mechanism. The pruning mechanism is configured with Pruning Threshold of 50\% and Toggle is set to engage task dropping reactively. 

Figure~\ref{fig:pruningBatch} shows that, for all heuristics under both constant and spiky arrival pattern, pruning mechanism improves the robustness. 
Pruning mechanism makes the largest impact for MSD and MMU. These heuristics attempt to map tasks with short deadlines and, thus, low chance of success. By limiting these heuristics to map tasks whose chance is beyond a certain threshold, their overall system robustness is improved.

\subsection{Impact of Pruning Mechanism on Homogeneous Systems}
In addition to mapping heuristics for heterogeneous system, we also conduct experiments on homogeneous mapping heuristics to evaluate the impact of pruning mechanism. Pruning configurations are set to use reactive Toggle and Pruning Threshold of 50\%.
 
Figure~\ref{fig:arrivalHomo} shows that, in all levels of oversubscription, applying pruning mechanism to homogeneous systems significantly increases system robustness (by up to 28\%) for all mapping heuristics on both constant and spiky arrival pattern. 
Importantly, as the oversubscription level increases, the impact of pruning mechanism is more substantial. With 25K tasks arrival rate, in constant arrival pattern, EDF and SJF can only achieve 4\% and 10\% robustness, respectively. Coupling pruning mechanism into these heuristics raises both the robustness to more than 30\%. The reason is that, similar to heterogeneous systems, pruning mechanism allows the system to avoid mapping unlikely-to-succeed tasks, which appear more often under higher levels of oversubscription. 

Based on the observations of this experiment, we can conclude that, pruning mechanism works equally as well and provides as much benefit to homogeneous systems as to the heterogeneous systems.

\section{Related Works}
\label{sec:rw}
Mapping tasks in HC systems has shown to be an NP-complete problem~\cite{coffman76,Ibarra77}. Multiple research works have been undertaken that try to maximize robustness of the HC systems. In this section, we review research works closely related to this study.

In~\cite{Shestak08}, Shestak \etal present the use of probability density functions (PDF) and mass functions (PMF) to capture uncertainty in task execution time. The authors use the robustness metric to showcase the effectiveness of the model using static resource allocation techniques. Their work lays out the method of convolving PMFs to estimate chance of success. We extend their approach to dynamically calculate task completion PMFs via convolution and to handle the case of probabilistic task pruning.

In an early work investigating heterogeneous scheduling~\cite{malone1983enterprise}, Malone \etal found out that lazy assignment, \ie deferring task scheduling to the latest possible, improves the performance of HC systems. They show that outperformance in comparison with eager assignment increases, both as heterogeneity increases and as error in run-time estimation increases. We use this lazy assignment as a foundation for probabilistic task deferring.

Malawski \etal evaluate dynamic scheduling of constrained tasks in clouds~\cite{malawski2015algorithms}. 
Their work concerns with the cost of provisioning homogeneous virtual machines (VMs) to process workflows, which is different to our focus on meeting deadlines of independent tasks on heterogeneous machines. The authors introduce workflow-aware algorithms that drop workflows result in the loss of high-priority tasks, to increase the performance. 

Khemka \etal investigate oversubscribed heterogeneous system resource management in~\cite{khemka2015utility}. They propose and investigate a parameterized method of utility function creation using priority, utility class, and urgency. They track execution times using an ETC matrix with deterministic scalar execution times, whereas we model the times via PMFs to accurately capture the uncertainty in task completion, especially in the presence of task-dropping. Tasks are dropped from their system after a task has reached a certain utility threshold, as opposed to our probabilistic deadline-focused approach.

Li \etal propose a task scheduling method for a live video streaming using cloud service~\cite{li2016vlsc}. Their video segment transcoding system consider tasks with a hard deadline and drops those that miss their deadline. However, there is no probabilistic task pruning in place. Although the mechanism described herein uses live video streaming as a motivation, the mechanism can be applied to any system with hard deadlines.

Delimitrou and Kozyrakis propose and explore Paragon~\cite{paragon13}, an immediate-mode scheduling system for heterogeneous systems. Historical task execution information are used by singular value decomposition to classify incoming tasks based on their heterogeneity and potential co-location interference. These classifications are used to select mappings based on interference and heterogeneity~\cite{delimitrou2014quality}. The mapping heuristics use scalar execution times to make decisions, whereas we focus on the uncertainty that is captured by modeling execution times via PMFs, and consider both immediate- and batch-mode heuristics. Another difference is in the performance metrics. Paragon tasks have no deadlines and its goal is to maximize speedup.

The act of proactive task dropping is similar to network packet queuing policy (\eg RED~\cite{patel2010performance}) in dropping network packets. However, there are key differences: (1) Our problem is dealing with big tasks that are less delay sensitive than network packets, therefore, more sophisticated consideration task deferring and dropping decision is allowed; (2) We prioritize task deferring that offers more chance to tasks to get scheduled later; (3) Our performance metric is the number of tasks meeting their deadlines, rather than latency or throughput. Hence, we believe this research is more closely related to HC systems but results can potentially be useful and applied to other domains as well. 

\section{Conclusion and Future works}
\label{sec:conclsn}
In this research, our goal was to enhance robustness of an HC system, deployed for serverless computing, via pruning task requests with low chance of success. We introduced a stand alone pruning mechanism that can be plugged into any task mapping heuristic without requiring any change in the resource allocation system. Evaluation results of applying the pruning mechanism on widely-used mapping heuristics in both homogeneous and heterogeneous systems showed that probabilistic dropping and deferring of unlikely-to-succeed tasks can increase the overall system robustness. The improvement was more remarkable (up to 35 percentage point) for heuristics with far-from-optimal mapping decisions (\eg MMU and EDF). Even in the case of MinMin mapping heuristic, the pruning mechanism led to 15 percentage point increase in robustness. We can conclude that, regardless of the underlying mapping heuristic, probabilistic task pruning can effectively improve robustness of serverless computing platforms, particularly when the system is oversubscribed. 

From the system perspective, we believe that, probabilistic task pruning improves energy efficiency by saving the computing power that is otherwise wasted to execute failing tasks. Such saving in computing can also reduce the incurred cost of using cloud resources for the serverless computing provider. In the future, we plan to measure such improvements in energy and incurred cost. Another future plan is to 
work on pruning methods that incorporate cost/priority of tasks, when considering dropping each individual task.

\section*{Acknowledgments}
We would like to thank the reviewers for their time and comments.
Portions of this research were conducted with high performance computational resources provided by the Louisiana Optical Network Infrastructure (http://www.loni.org) \cite{LONI}.
This research was supported by the Louisiana Board of Regents under grant number LEQSF(2016-19)-RD-A-25. 

\bibliographystyle{IEEEtran}
\balance
\bibliography{references}
\end{document}